\documentclass[aps,twocolumn,groupeaddress,prb]{revtex4}

\usepackage{amsfonts, amsmath, amssymb,latexsym}
\usepackage{subfigure}
\usepackage{graphicx}

\newcommand{\me}{\mathrm{e}}

\newcommand{\dif}{\mathrm{d}}

\begin{document}

\title{Generalizations of the Fuoss Approximation for Ion Pairing}

\author{P. Zhu}
\email{zpeixi@tulane.edu}
\author{X. You}
\email{xyou@tulane.edu}
\author{L. R. Pratt} 
\email{lpratt@tulane.edu}
\author{K. D. Papadopoulos}
\email{kyriakos@tulane.edu}
\affiliation{Department of Chemical and Biomolecular Engineering, Tulane
University, New Orleans, LA 70118}

\date{\today}

\begin{abstract} 
An elementary statistical observation identifies generalizations of the
Fuoss approximation for the probability distribution function that
describes ion clustering in electrolyte solutions. The simplest
generalization, equivalent to a Poisson distribution model for
inner-shell occupancy, exploits measurable inter-ionic correlation
functions, and  is correct at the closest pair distances whether
primitive electrolyte solutions models or molecularly detailed models
are considered, and for low electrolyte concentrations in all cases. 
With detailed models these generalizations  includes non-ionic
interactions and solvation effects. These generalizations are relevant
for computational analysis of bi-molecular reactive processes in
solution. Comparisons with direct numerical simulation results show that
the simplest generalization is accurate for a slightly supersaturated
solution of tetraethylammonium tetrafluoroborate in propylene carbonate
([tea][BF$_4$]/PC), and also for a primitive model associated with the
[tea][BF$_4$]/PC results. For [tea][BF$_4$]/PC, the atomically detailed
results identify solvent-separated nearest-neighbor ion-pairs. This
generalization is examined  also for the ionic liquid
1-butyl-3-methylimidazolium tetrafluoroborate ([bmim][BF$_4$]) where the
simplest implementation  is less accurate.  In this more challenging
situation an augmented maximum entropy procedure is satisfactory, and
explains the more varied near-neighbor distributions observed in that
case.
\end{abstract}

\maketitle 

\section{Introduction} 
Ion clustering has long been an essential ingredient of our physical
understanding of electrolyte solutions at elevated concentrations.
\cite{RS,Bjerrum,Fuoss:1934p14292,Fowler-Guggenheim,%
Reiss:1956p14294,Friedman:1961p14569,StillingerJr:1968p14285,%
Given:1997p14295,%
Camp:1999p14354,ISI:000081550900132,%
ISI:000232442500050} To describe pairing of a counter-ion of type
$\gamma$ with an ion of type $\alpha$, we focus on
the radial distribution of the \emph{closest}
$\gamma$-ion to a distinguished $\alpha$-ion.   We denote that
normalized radial distribution by $g_{\gamma\vert\alpha}^{(1)}(r)$.  A
famous discussion of Fuoss \cite{Fuoss:1934p14292} arrived at the
approximation
\begin{eqnarray}
\ln g_{\gamma\vert\alpha}^{(1)}(r) \approx \frac{\beta q^2}{\epsilon r}  -
4\pi\rho\int_{d_{\gamma\alpha}}^r \me^{\frac{\beta q^2}{\epsilon x} }x^2\dif x~, r \ge d_{\gamma\alpha}
\label{eq:Fapprox}
\end{eqnarray}
for a primitive model of 1-1 electrolyte as in
FIG.~1. Here $q$ is the magnitude of the formal ionic
charges, $d_{\gamma\alpha}$ is the distance of closest approach,
$\epsilon$ is the solution dielectric constant, $2\rho$ is number
density of ions, and $\left(k\beta\right)^{-1}=T$ is the temperature. We
propose and test generalizations of Eq.~\eqref{eq:Fapprox} in the
following.

Several complications of the distributions of near-neighbor ion-pairs
motivate the generalizations that we develop.  Firstly, ion-clustering
can be particularly sensitive to non-ionic interactions. Comparison 
(FIG.~1) of
atomically-detailed simulation results \cite{Yang:2009p11328,Yang:2010p14158}
with those of a corresponding
primitive model \cite{Towhee}  straightforwardly
exemplifies that point. Eq.~\eqref{eq:Fapprox}
only treats classic ionic interactions.  Secondly, even for primitive models the Fuoss
approximation can be unsatisfactory (FIG.~2).
Thirdly, nearest-neighbor distributions generally depend on
which ion of an ion-pair is regarded as the central  ion
(FIG.~3). The radial distribution of the anion nearest to a cation
is different from the radial distribution of the  cation nearest to an anion,
$g_{\mathrm{\alpha\vert\gamma}}^{(1)}(r) \ne
g_{\mathrm{\gamma\vert\alpha}}^{(1)}(r)$.  The approximation Eq.~\eqref{eq:Fapprox} is symmetric
$g_{\mathrm{\alpha\vert\gamma}}^{(1)}(r) = 
g_{\mathrm{\gamma\vert\alpha}}^{(1)}(r)$

We are lead then to generalizations by recalling that the
probability that a ball of radius $r$ centered on an $\alpha$-ion is
\emph{empty} of $\gamma$-ions can be obtained from
\begin{eqnarray}
p_{\gamma\vert\alpha}(n=0) =  4 \pi  \rho_\gamma
\int\limits_r^\infty g_{\gamma\vert\alpha}^{(1)}(x)  x^2 \dif x ~,
\label{eq:p0}
\end{eqnarray}
the assessment of the probability that the \emph{nearest} $\gamma$-ion
is  further away than $r$.  The simple estimate 
\begin{eqnarray}
p_{\gamma\vert\alpha}(n=0) \approx \exp\left\lbrack
-\left\langle n_{\gamma\vert\alpha}\left(r\right)\right\rangle\right\rbrack~,
\label{eq:PoissonApprox}
\end{eqnarray}
 with $\left\langle
n_{\gamma\vert\alpha}\left(r\right)\right\rangle = 4 \pi  \rho_\gamma \int^r_0 
g_{\gamma\alpha}(x) x^2\dif x,$ $\rho_\gamma$ the density
of $\gamma$ ions, and $ g_{\gamma\alpha}(x)$ the
conventional radial distribution function, follows from 
the assumption of the Poisson distribution for that
probability. Evaluating the derivative of
Eq.~\eqref{eq:p0} using Eq.~\eqref{eq:PoissonApprox} gives
\begin{eqnarray}
g_{\mathrm{\gamma\vert\alpha}}^{(1)}(r) \approx
g_{\mathrm{\alpha\gamma}}(r) \exp\left\lbrack - 4\pi\rho_\gamma\int_0^r 
 g_{\gamma\alpha}(x)  x^2 \dif x\right\rbrack~.
\label{eq:GFuoss}
\end{eqnarray}
For $g_{\mathrm{\alpha\gamma}}(r) = 1$ (no correlations), this is the
Hertz distribution that is correct for that
case.\cite{MAZURS:Neiprd,Chandrasekhar} We recover the Fuoss
approximation with $\ln g_{\mathrm{\alpha\gamma}}(r)
\approx -\beta q_\alpha q_\gamma /\epsilon r= \beta q^2/\epsilon r$ for
$r> d_{\alpha\gamma}$, and zero (0) otherwise.   This derivation of the
Fuoss approximation Eq.~\eqref{eq:Fapprox} seems not to have been given
before.   Nevertheless, the suggested approximation
Eq.~\eqref{eq:GFuoss} is a standard idea in the context of
scaled-particle theories of the hard-sphere fluid.\cite{MAZURS:Neiprd}

\begin{figure}[h]
    \includegraphics[width=2.5in]{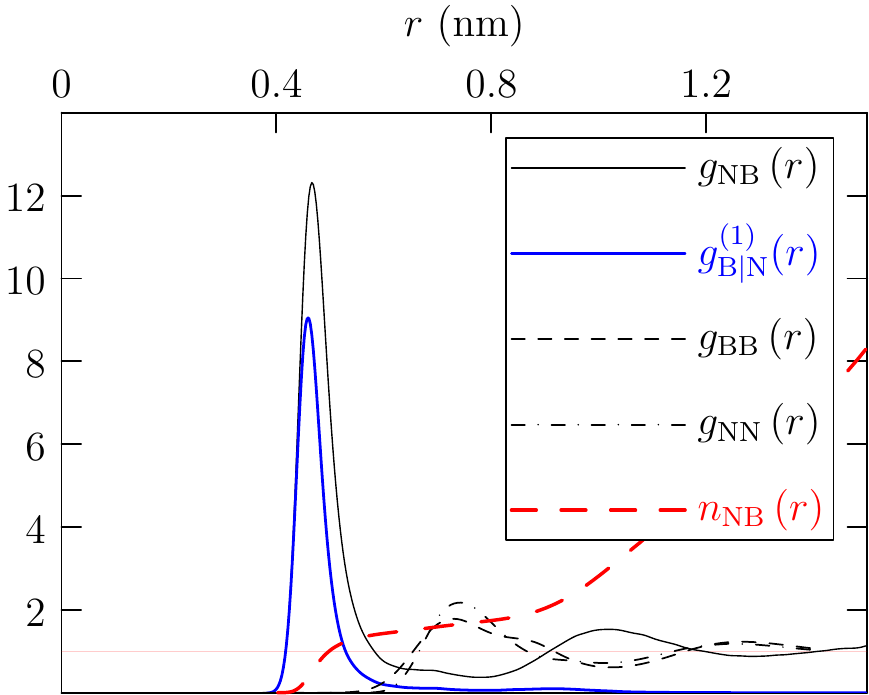}
    \includegraphics[width=2.5in]{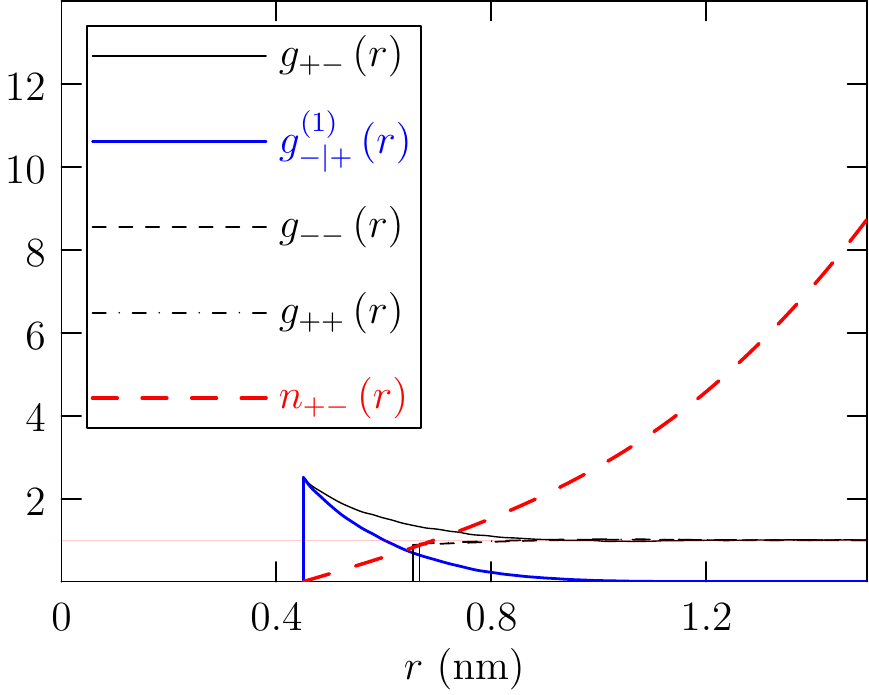}
\caption{ Upper-panel: Ion-ion radial distribution functions for
atomically-detailed simulation \cite{Yang:2009p11328,Yang:2010p14158} of
tetraethylammonium tetrafluoroborate in propylene carbonate
([tea][BF$_4$]/PC) at $T$ = 300~K, $p$ = 1~atm, and the slightly
supersaturated concentration of 1 mol/dm$^3$. $g_{\mathrm{B\vert
N}}^{(1)}(r)$ is the radial distribution of the nearest B-neighbor of an
N-atom.  Lower panel: Results for a corresponding primitive model with
dielectric constant and with ion charges and sizes matched to the
[tea][BF$_4$]/PC case above.  Specifically the model dielectric constant
is $\epsilon = 60$, and $d_{++} = 0.6668$~nm, $d_{--} = 0.6543$~nm,
$d_{-+} = 0.45$~nm are distances of closest approach for the hard
spherical ions. The lower panel was produced by Monte Carlo simulation of
a neutral system of 2$\times$500 hard spherical ions in conventional
cubical periodic boundary conditions at the same temperature and
concentration as the results above, utilizing the Towhee\cite{Towhee}
package adapted to the present system.\label{fig:TEABF4_rdf}}
\end{figure}

As discussed below, the Poisson result Eq.~\eqref{eq:PoissonApprox} follows from a maximum
entropy development when the information supplied is the expected
occupancy of the inner-shell.\cite{HummerG:Anitm,PrattLR:Thehea,PrattLR:Molthe} That
information is sufficient if the occupancy $n\left(r\right)$ is always
low, \emph{i.e.}, rarely larger than one. Thus, in contrast to the Fuoss
approximation, Eq.~\eqref{eq:GFuoss} is  correct for small $r$ because
the expected coordination number tends to zero then. For
the same reason, the Poisson approximation Eq.~\eqref{eq:PoissonApprox}
is correct at low electrolyte concentration, and even when the
solvent is treated at atomic resolution.
Furthermore, it is
natural to guess cation-anion chain or ring structures when ionic
interactions drive well developed clustering. FIG.~1
shows a mean coordination number of less than two for counter-ion
neighbors closer than about 0.5~nm, and supports the chain/ring picture
of ion clusters formed. It is plausible therefore that a choice of
inner-shell radii leading to small coordination numbers should validly
describe important features of well-developed ion-clustering.  

For computational analysis of reactive bi-molecular encounters in solution,
identification of geometries of closest molecular pairs is
critical.\cite{Chempath:2008p368,Chempath:2010p15651} Because
it is correct for low concentration and for small $r$ in any case, Eq.~\eqref{eq:GFuoss} should be
regarded as the general resolution of those questions.

When coordination numbers exceed one with reasonable probability,
information on the expected number of pairs of counter-ions in the
inner-shell should improve a maximum entropy model of these
probabilities.\cite{HummerG:Anitm,PrattLR:Thehea,PrattLR:Molthe}  A maximum entropy
model involving pair information would predict the
$g_{\mathrm{\alpha\vert\gamma}}^{(1)}(r) \ne
g_{\mathrm{\gamma\vert\alpha}}^{(1)}(r)$ asymmetry. For a 1-1
electrolyte, the generalization Eq.~\eqref{eq:GFuoss} is symmetrical in
accord with the Fuoss approximation. The extent to which the observed
asymmetry is significant gives an indication whether the Poisson
approximation is adequate.

\begin{figure}
    \includegraphics[width=2.5in]{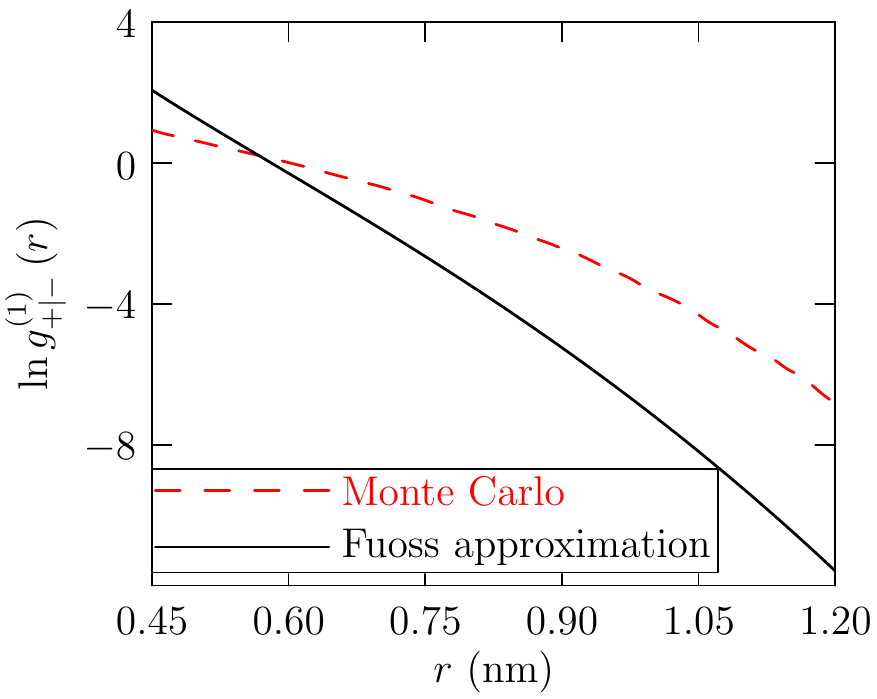} 
    \caption{ Comparison of the Fuoss approximation for
    $g_{\mathrm{+\vert -}}^{(1)}(r)$ to Monte Carlo results of
    FIG.~1.
    \label{fig:MC_Fuoss_Primitive}}
\end{figure}

\begin{figure}
    \includegraphics[width=2.5in]{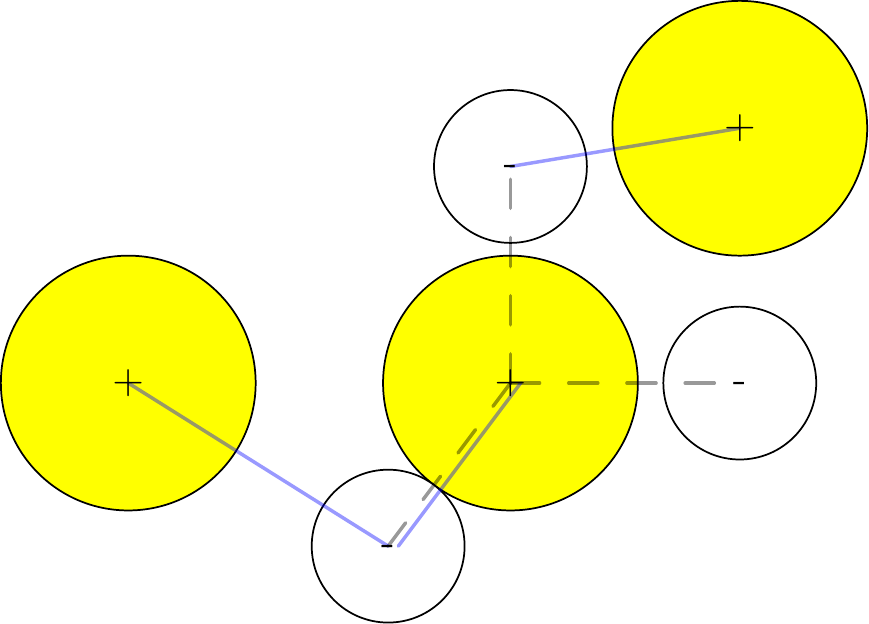}
\caption{An example showing that the distribution of  anions nearest to
a cation is generally different from the distribution of the cations
nearest to an anion. The dashed lines indicate the  nearest distances of
a cation to each of the three anions.   The solid lines mark the nearest
distances of a anion to each of the three cations.\label{fig:balls}}
\end{figure}

In this work, the Poisson approximation (Eq.~\eqref{eq:GFuoss}) is
tested using three distinct simulation data sets.   Two of these data
sets have been noted already in considering FIG.~1.
Those calculations treated solutions of tetraethylammonium
tetrafluoroborate in propylene carbonate, one at atomic resolution
([tea][BF$_4$]/PC) and the other on the basis of a primitive electrolyte solution
model over a range of concentrations.  The third data set treated the
ionic liquid 1-butyl-3-methylimidazolium tetrafluoroborate
([bmim][BF$_4$]). To ensure the correct correspondence of the necessary
simulation details with the results as they are discussed, those details
are provided in the captions of the figures providing the simulation
results.

\section{Results and Discussion}

For [tea][BF$_4$]/PC, comparison (FIG.~4) of the
numerical data with the approximation Eq.~\eqref{eq:GFuoss} shows
agreement over a distance range wider than the sizes of the molecules as
judged by the radial distributions (FIG.~1). These
near-neighbor distributions show bi-modal probability densities with
maxima at $r\approx$ 0.5~nm and 0.9~nm.  These correspond, respectively,
to a contact ion pair and to a solvent-separated  near-neighbor
ion-pair.  Thus the Poisson approximation Eq.~\eqref{eq:GFuoss} in this
case includes solvation structure in characterizing inter-ionic
neighborship.

\begin{figure}[h]
	\begin{center}
    \includegraphics[width=2.5in]{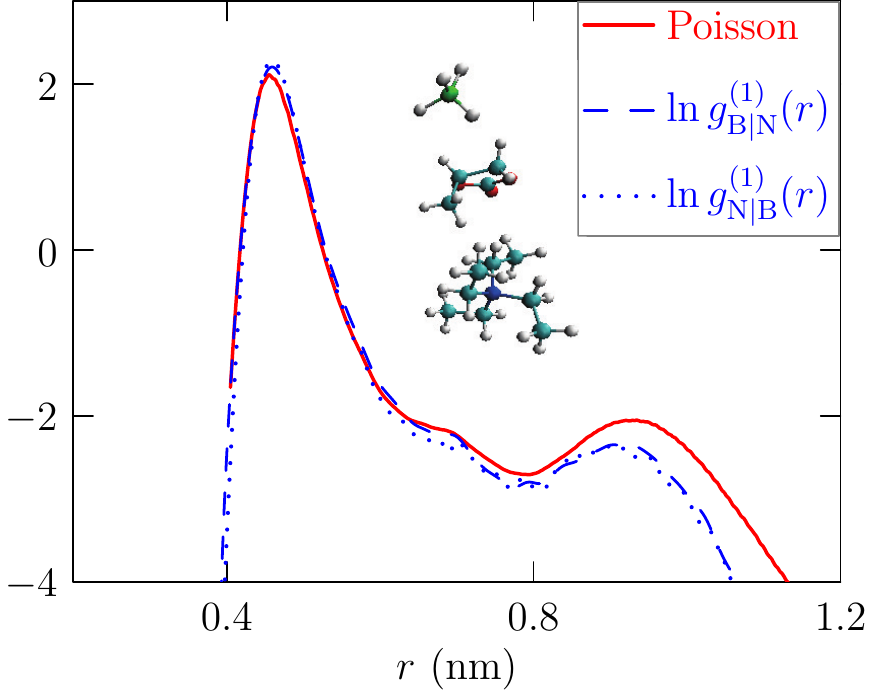}
    \end{center}
\caption{For the [tea][BF$_4$]/PC case of FIG.~1,
comparison of the numerical data with the approximation
Eq.~\eqref{eq:GFuoss}. The local maximum at $r\approx$ 0.9~nm identifies
solvent-separated nearest-neighbor ion-pairs.   In this case, asymmetry
of the two observed near-neighbor distributions is slight.  That
suggests that more elaborate maximum entropy models are unnecessary, and
indeed the Poisson approximation is accurate.  The embedded molecular
graphic shows one of the solvent-separated nearest-neighbor
BF$_4{}^-$\ldots PC \ldots tea$^+$ structures observed.
\label{fig:g1PoissonTEA}}
\end{figure}

\begin{figure}[h]
	\begin{center}
    \includegraphics[width=2.5in]{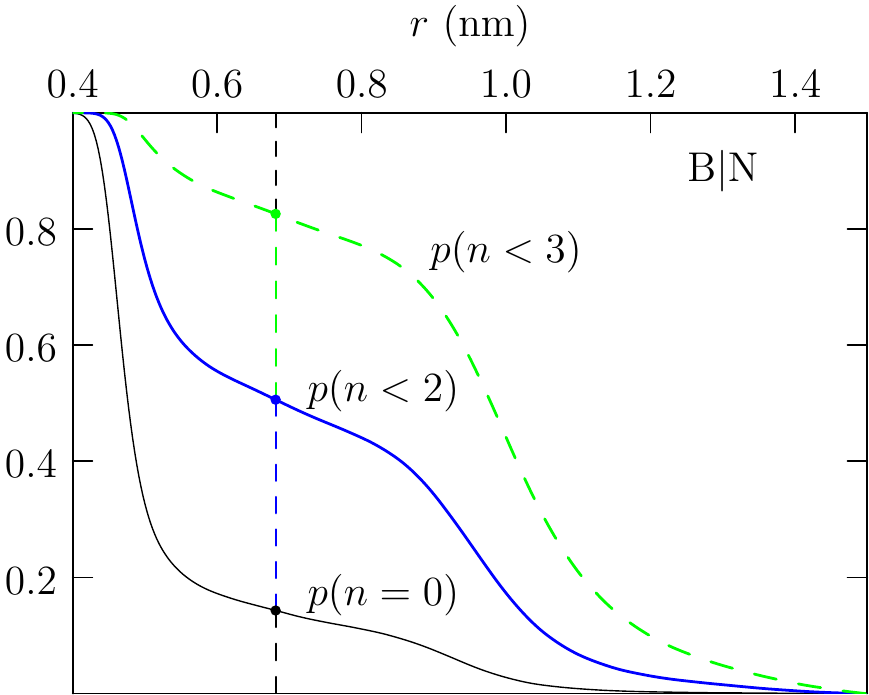}
    \includegraphics[width=2.5in]{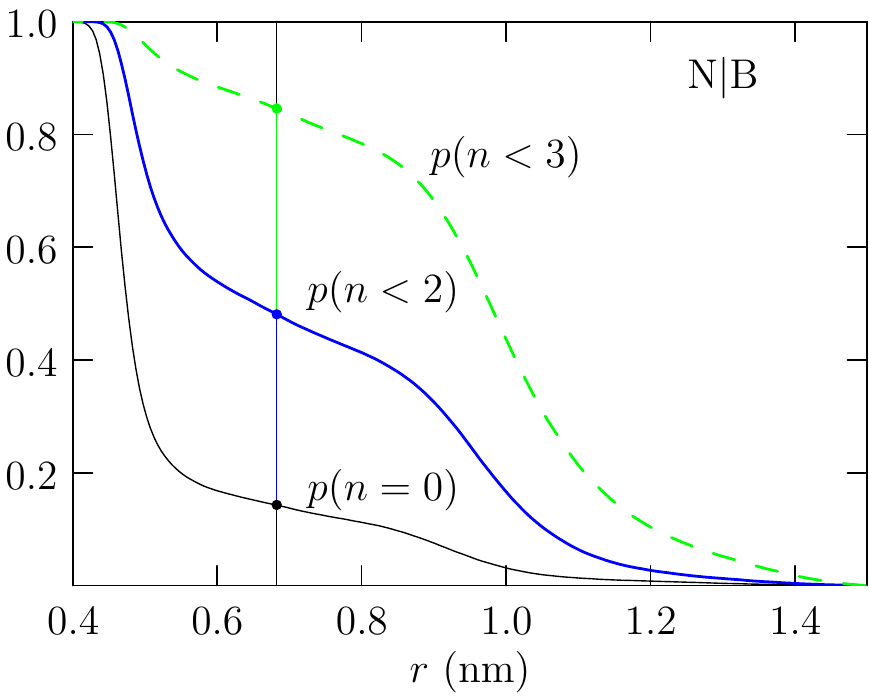}
    \end{center}
\caption{Occupancy probabilities as functions of the observation sphere
radii $r$ for the [tea][BF$_4$]/PC case of FIG.~1.  Upper
panel:   probabilities for occupancy by the B-atom of the BF$_4{}^-$
anion of the inner-sphere of the N-atom of tetraethylammonium cation
(tea$^+$). Lower panel: probabilities for occupancy by the N-atom of the
cation of the inner-sphere of the B-atom of the anion.  The
curves  lowest in each panel are similar, showing symmetry
displayed also in FIG.~4. \label{fig:Probability_TEABF4}}
\end{figure}

A plateau between $r\approx$ 0.5~nm and 0.9~nm in occupancy
probabilities (FIG.~5) indicates saturation
of counter-ion probability, and marks the inter-shell region. At the
distance $r$ indicated by the vertical line, the coordination numbers
$n$ = 1, 2 predominate, supporting the idea of the formation of
cation-anion chain and ring structures. The two sets of probabilities
(FIG.~5) are qualitatively similar,
reinforcing the symmetry of FIG.~4.

\begin{figure}
	\begin{center}
    \includegraphics[width=2.5in]{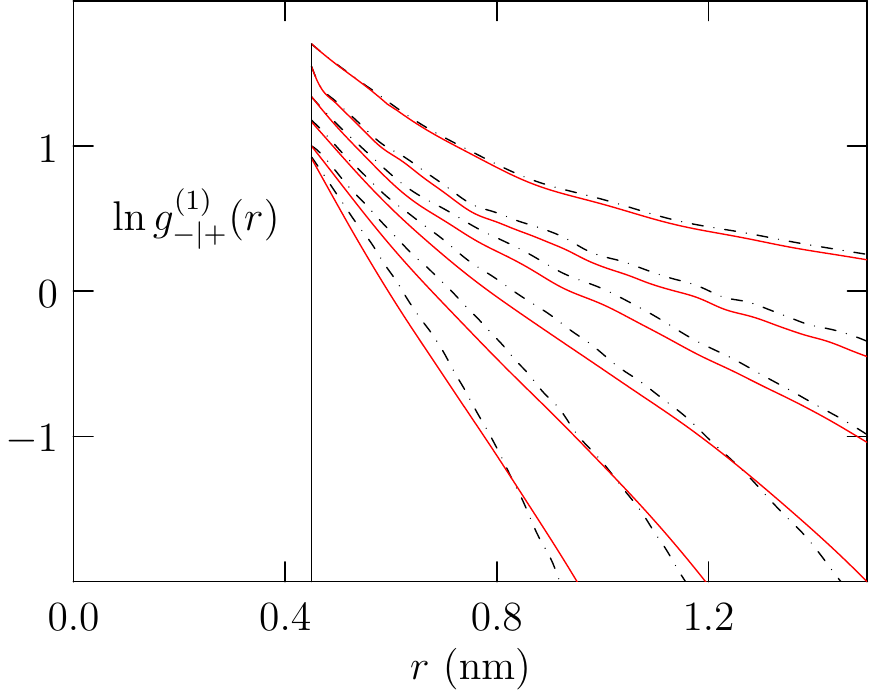}
    \end{center}
\caption{Dashed curves are Monte Carlo results for the primitive model
of FIG.~1 and the solid lines are the Poisson
approximation, Eq.~\eqref{eq:GFuoss}.  From top to bottom, the distinct
cases correspond to concentrations 0.01, 0.05, 0.10, 0.20, 0.40, and
0.80~mol/dm$^3$, $T$ = 300K for each case.  For the highest
concentration, the system size is 2$\times$400 ions.   For all other
cases, the system size is 2$\times$200 ions. At the lowest
concentration here the distribution of the nearest neighbor
$g_{-|+}^{(1)}(r)$ is close to the full radial distribution function
$g_{+-}(r)$. \label{fig:MC_Poisson_Primitive}}
\end{figure}

Results (FIG.~6) for the primitive model of
FIG.~1 examine the sufficiency  of the Poisson
approximation over a broader concentration range for such models. The
nearest-neighbor distributions are unimodal in this case. Correct at
small $r$ where the probability densities are highest and properly
normalized, the Poisson approximation Eq.~\eqref{eq:GFuoss} is accurate
over the whole range shown.

\begin{figure}[h]
   \includegraphics[width=3.0in]{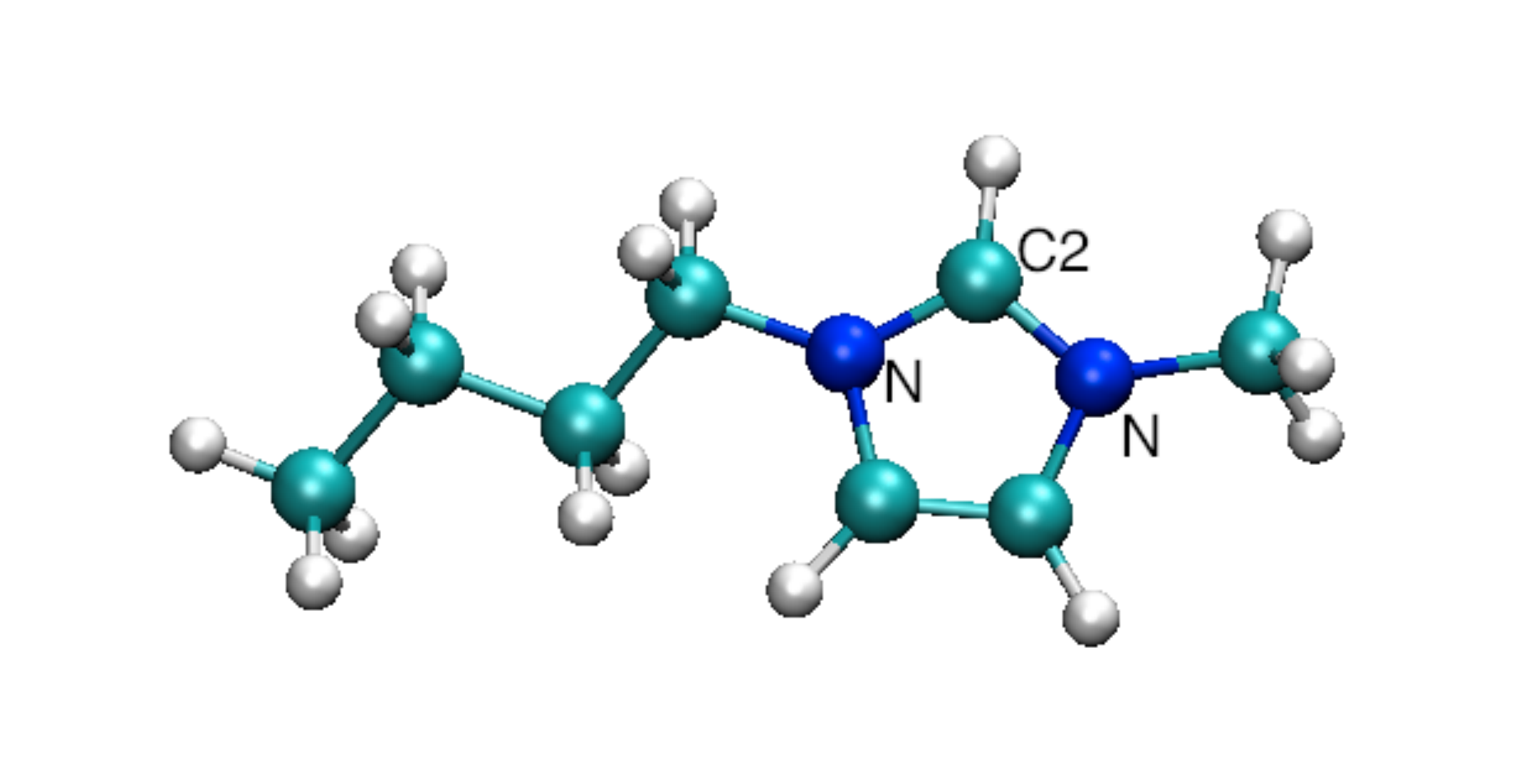}\newline
   \includegraphics[width=0.7in]{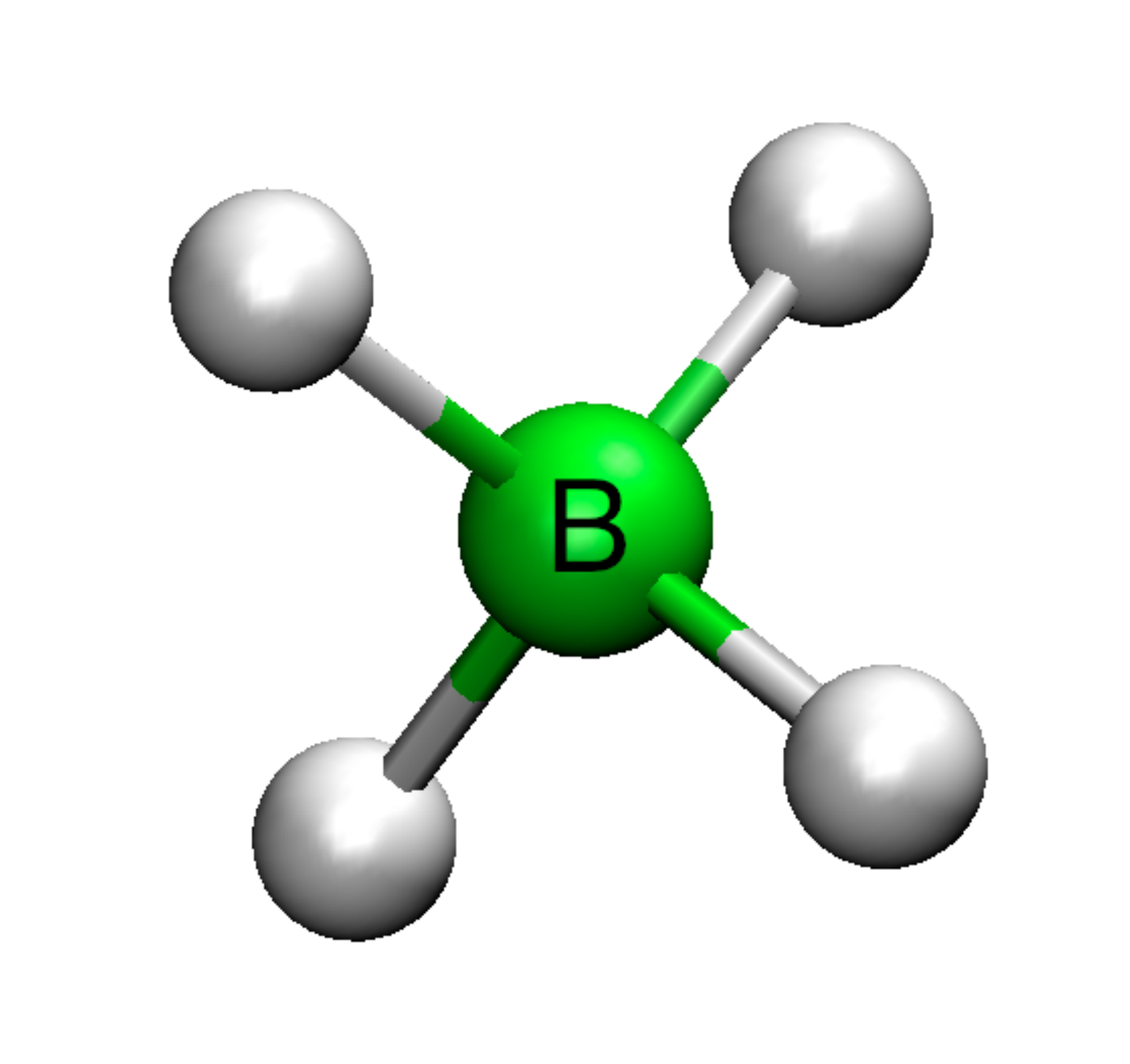}
\caption{Structures and atom labeling of cation
1-butyl-3-methylimidazolium (bmim$^+$) (top) and anion tetrafluoroborate
(BF$_4{}^-$). On the cation molecule, white, turquoise and dark blue
balls stand for hydrogen, carbon, and nitrogen atoms, respectively. On
the anion, white and green balls stand for fluorine and boron atoms,
respectively. \label{fig:BMIMBF4_structure}}
\end{figure}

\begin{figure}[h]
	\begin{center}
    \includegraphics[width=2.5in]{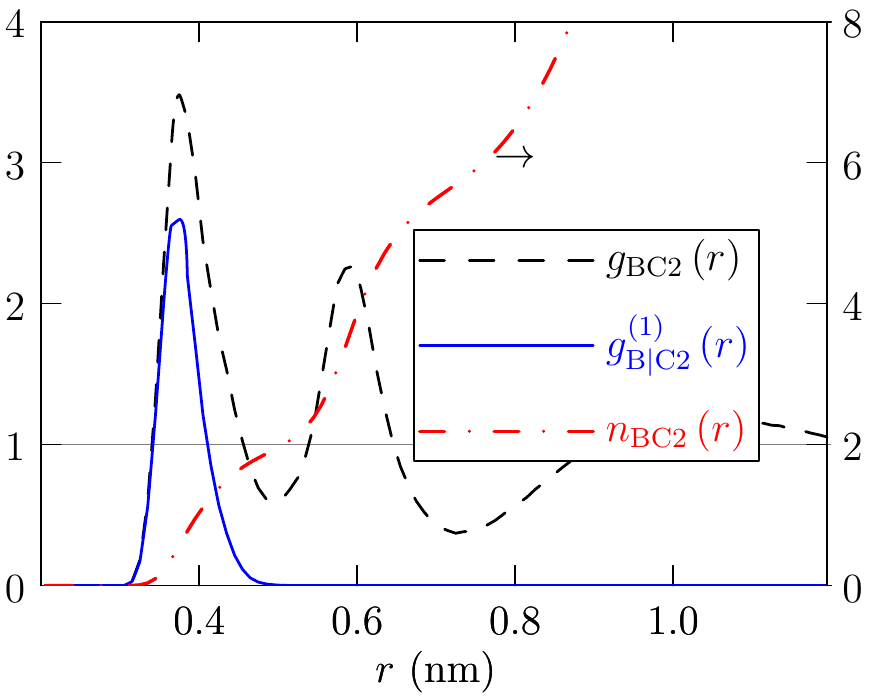}
    \end{center}
\caption{Radial distribution function of C2 atom on bmim$^+$ cation and
boron on BF$_4{}^-$ anion obtained from MD simulation. Force field
parameters and partial charge of the atoms were taken from Andrade
\emph{et al}.\cite{de_andrade_computational_2002} The initial unit cell,
with a dimension of 40~$\times$~40~$\times$~40 {\AA} is uniformly packed
with 190 ion pairs using Packmol.\cite{martnez_packmol:package_2009} MD
simulation was performed using AMBER10 at constant pressure (1 bar) and
temperature (298.5K). The system was first minimized, followed by
0.2ns equilibrium at a time step of 0.2~fs, then 1.3~ns equilibrium at a
time step of 2~fs. Radial distribution functions were extracted from a
production run of 3.0~ns.
\label{fig:BMIMBF4_rdf}}
\end{figure}

\begin{figure}[h]
	\begin{center}
    \includegraphics[width=2.5in]{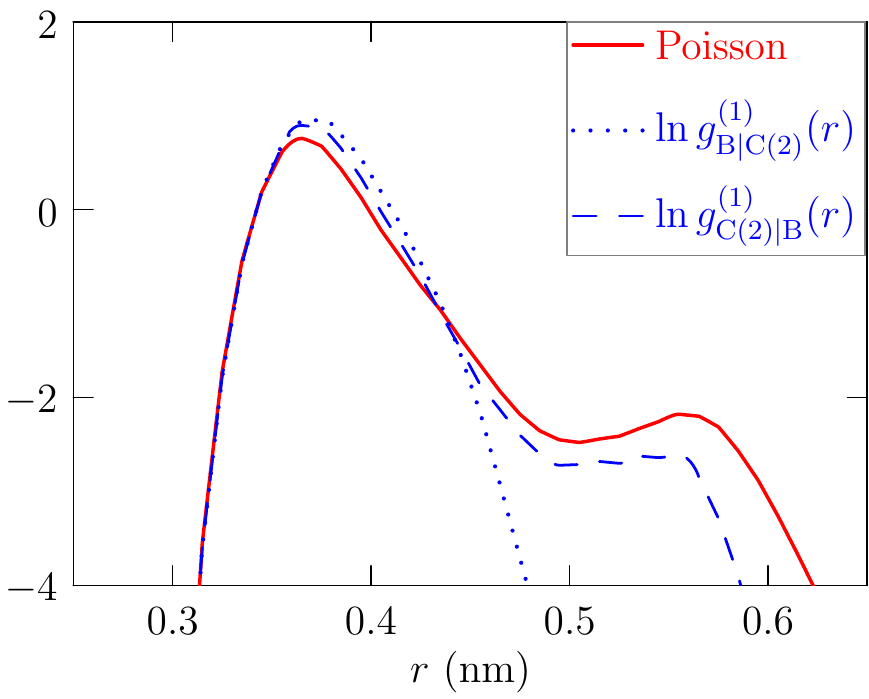}
    \end{center}
\caption{Nearest-neighbor distributions for the ionic liquid case of
FIG.~8.  In this case with no solvent, the asymmetry
of the two observed near-neighbor distributions is marked.  An
outer-sphere nearest-neighbor is exhibited in one case but not the other case
where the Poisson approximation Eq.~\eqref{eq:GFuoss} is inaccurate.
\label{fig:g1Poisson_BMIMBF4}}
\end{figure}

Another example is the ionic liquid [bmim][BF$_4$], with molecular structure shown in
FIG.~7 and radial distribution functions in
FIG.~8.  The Poisson approximation
(Eq.~\eqref{eq:GFuoss}) agrees with the observed $g_{\mathrm{C2\vert
B}}^{(1)}(r)$ at short range and displays a second maximum
characterizing non-contact nearest neighbors, though in this case there
is no additional solvent. The near-neighbor B$|$C2 distribution
(FIG.~9), on the other hand, lacks a second
maximum. Thus $g_{\mathrm{C2\vert B}}^{(1)}(r)$ and $g_{\mathrm{B\vert
C2}}^{(1)}(r)$ for ionic liquid [bmim][BF$_4$] display the generally
expected asymmetry. This asymmetry is also reflected  in occupancy
probability profiles (FIG.~10). More general
theoretical models are required for such cases, and we return to
that theoretical discussion now.

\begin{figure}[h]
	\begin{center}
    \includegraphics[width=2.5in]{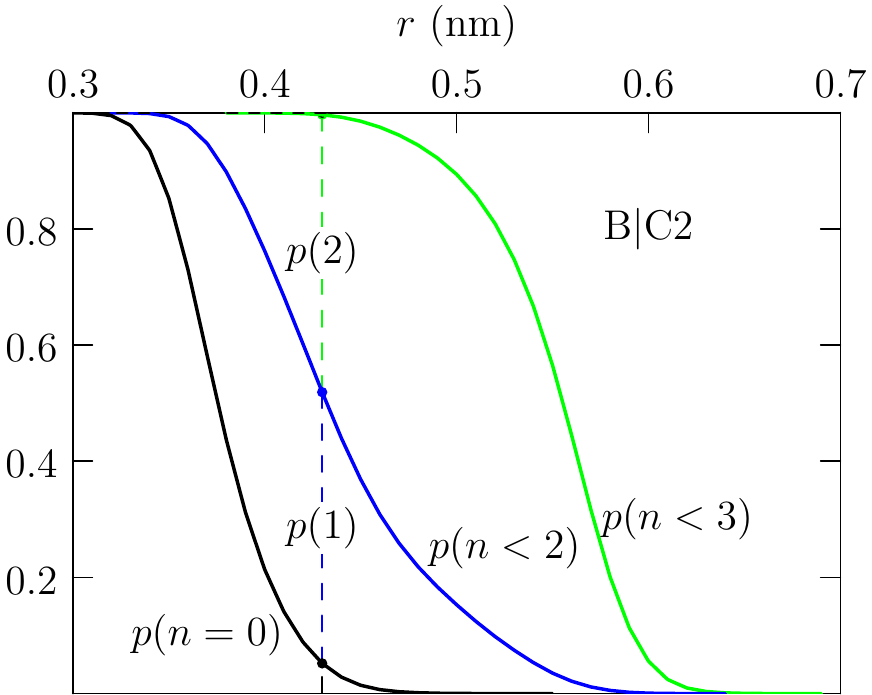}
    \includegraphics[width=2.5in]{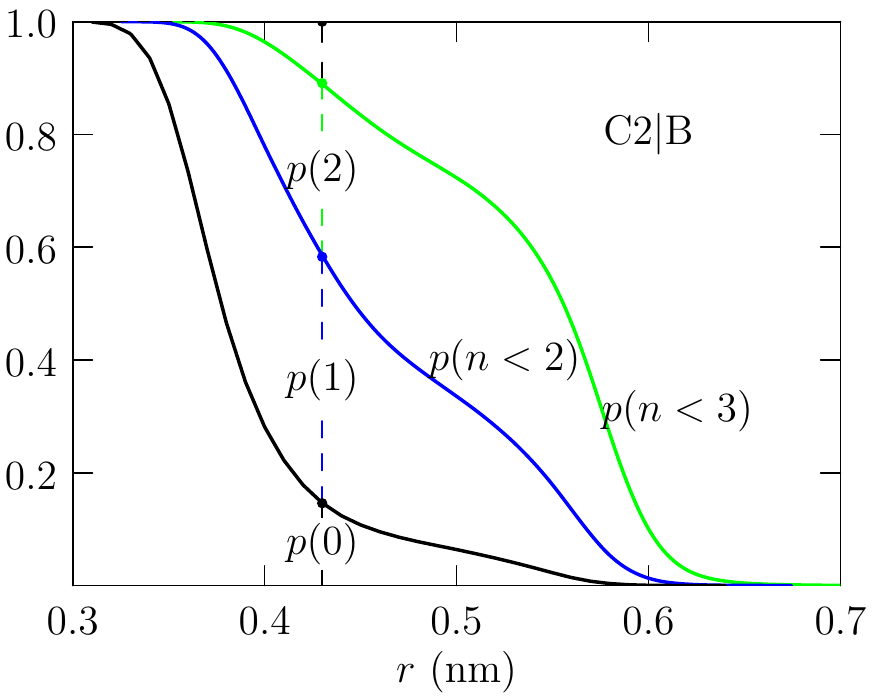}
    \end{center} 
    \caption{Occupancy probabilities as functions of the observation
    sphere  radii $r$ for [bmim][BF$_4$] (FIG.~8).  Upper
    panel:   probabilities for occupancy by the B-atom of the BF$_4{}^-$
    anion of the inner-sphere of the mid-C-atom of
    1-butyl-3-methylimidazolium (bmim$^+$) cation (labeled as C2 in
    FIG.~7).  Lower panel: probabilities for
    occupancy by the mid-C-atom of the cation of the inner-sphere of the
    B-atom of the anion.  These results illustrate the possibility of a
    structural motif of ion clusters as chains and rings,  \emph{i.e.} 
    at the $r$ = 0.43~nm distance of the vertical dashed line
    probabilities of 1 and 2 neighbors predominate. 
    They also demonstrate asymmetry of the distributions of near-neighbor
    distances in their dependence on choice of the central ion, \emph{i.e.}
    the distribution of nearest anions to a cation is different from the
    distribution of the nearest cations to an anion. Since $p(3)$ is
    larger in the lower panel than in the upper panel, the BF$_4{}^-$
    anion is more likely to be a three-way junction in this analysis
    than is the bmim$^+$ cation.
\label{fig:Probability_BMIMBF4}}
\end{figure}

\section{Maximum Entropy Modeling}

The Poisson distribution $\hat{p}\left(n\right)= \left\langle
n\right\rangle^n\me^{-\left\langle n\right\rangle}/{n!}$ describes
random occupancy consistent with the information $\left\langle
n\right\rangle=\left\langle n(r)\right\rangle$. Considering the relative
entropy,
\begin{eqnarray}
\eta(\{p\left(n\right)\})=~-\sum_{n\ge 0}p\left(n\right)\ln
\left(\frac{p\left(n\right)}{\hat{p}\left(n\right)}\right)~,
\label{eq:Cross_entropy}
\end{eqnarray}
the Poisson distribution is a maximum entropy distribution satisfying
the specific expected occupancy. If we have more information,
\emph{e.g.,} the binomial
moments\cite{HummerG:Anitm,PrattLR:Thehea,PrattLR:Molthe}
\begin{eqnarray}
\left\langle\binom{n}{j}\right\rangle=
\sum_{n\ge 0}p\left(n\right)\left(\frac{n!}{(n-j)!j!}\right)~,
\label{eq:constraint}
\end{eqnarray}
we can seek the distribution which maximizes
$\eta(\{p\left(n\right)\})$ and satisfies the broader set of
information.
 
With the binomial moments (Eq.~\eqref{eq:constraint}), the Poisson
distribution is seen to be correct if realized values of $n$ are rarely
bigger than one (1). If $n$ is never 2 or larger, binomial moments  $j\ge 2$
vanish.  When $j\ge 2$ binomial moments are small,
and that is consistent with Poisson prediction that they are zero. This
underlies our observation above the the Poisson model, $p\left( 0\right)
\approx \me^{-\left\langle n\right\rangle}$ of
Eq.~\eqref{eq:PoissonApprox}, is correct for small $\lambda$.

Beyond the mean occupancy, the next level of information is the
pair-correlation information $\left\langle
n(r)\left(n(r)-1\right)/2\right\rangle$, the expected number of pairs of
counter-ions in the indicated inner-shell. Carrying-out the maximization
for the case that pair information is available induces the model $p(n)
\propto \exp\left\lbrack -\lambda_1 n - \lambda_2
n(n-1)/2\right\rbrack/n!$, where $\lambda_1$, and $\lambda_2$ are
Lagrange multipliers adjusted to reproduce the information $\left\langle
n\right\rangle$ and $\left\langle n\left(n-1\right)/2\right\rangle$.
Explicitly addressing the normalization of these probabilities leads to
\begin{eqnarray}
p\left(n\right)=
\frac{
\left(\frac{1}{n!}\right)\me^{-\lambda_1 n-\lambda_2 n(n-1)/2}
}{
1 + \sum\limits_{m\ge 1}\left(\frac{1}{m!}\right)
\me^{-\lambda_1 m-\lambda_2 m(m-1)/2}
}~,
\label{eq:Lagrange}
\end{eqnarray}
and
\begin{eqnarray}
\ln p(0) = -\ln\left\lbrack 1 + \sum\limits_{n=1}^\infty
\left(\frac{1}{n!}\right)\me^{ -\lambda_1 n - \lambda_2
n(n-1)/2}\right\rbrack~.
\label{eq:lnp0}
\end{eqnarray}
$p(0)$ involves only the denominator of Eq.~\eqref{eq:Lagrange},
and can be considered a partition function sum over occupancy
states with $n$-dependent interactions and interaction strengths
adjusted to satisfy the available information.
The information required (FIG.~11) for this augmented
maximum-entropy model is only subtly different for the two cases.
Nevertheless, the results (FIG.~12) agree nicely with the
observed asymmetry.

\begin{figure}[h]
	\begin{center}
    \includegraphics[width=2.5in]{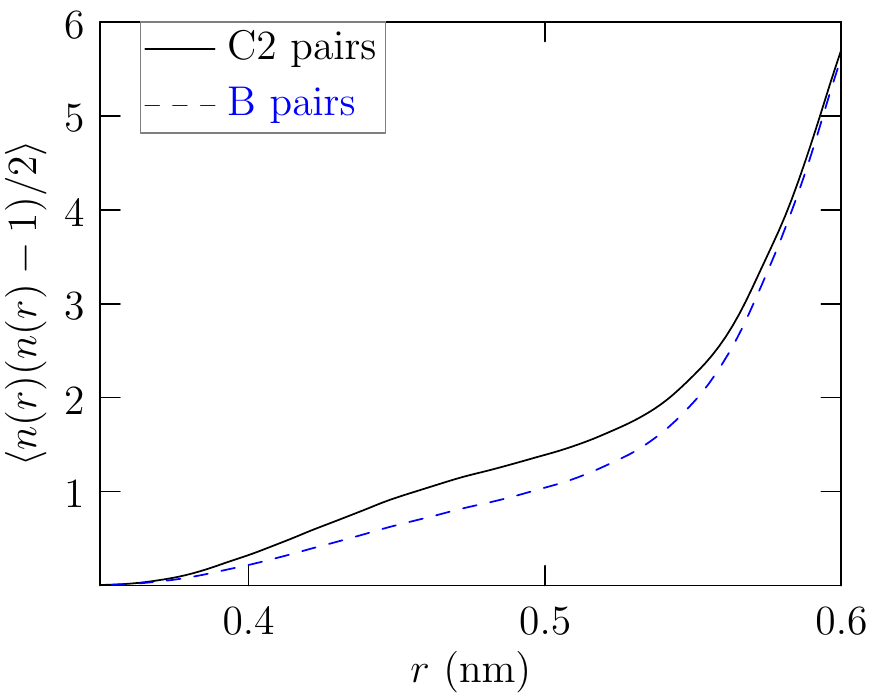}
    \end{center} 
    \caption{Pair information required for the augmented maximum entropy
    prediction of Eq.~\eqref{eq:lnp0}.  Solid curve: the number of C2
    pairs occupying a sphere of radius $r$ on a B atom. Dashed curve: 
    the number of B pairs occupying a sphere of radius $r$ on a C2 atom.
\label{fig:PairInfo}}
\end{figure}

\begin{figure}[h]
	\begin{center}
    \includegraphics[width=2.5in]{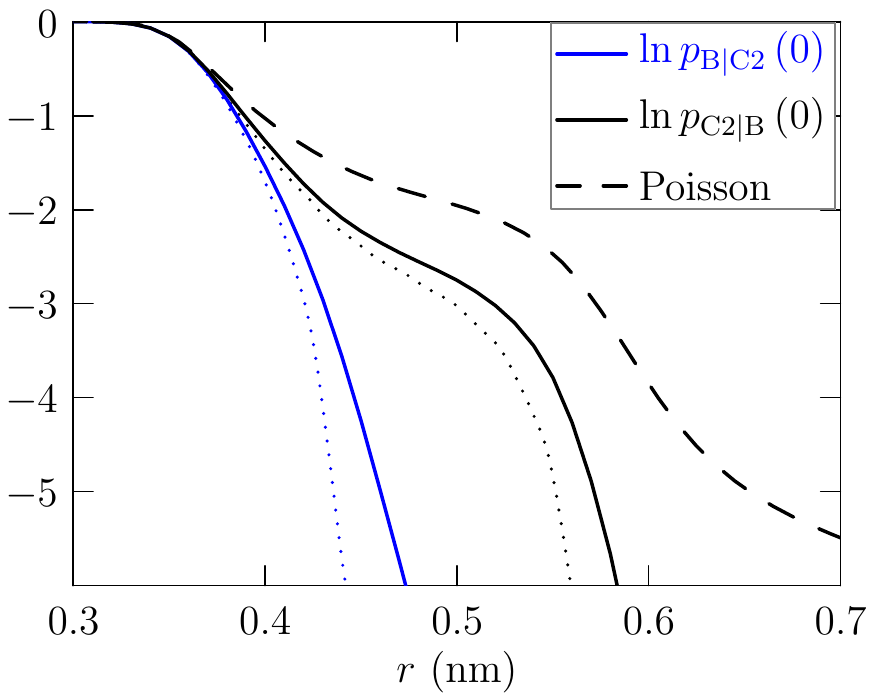}
    \end{center} 
    \caption{Analysis of the asymmetric $n=0$ probabilities of FIG.~10. 
     Solid curves: direct numerical simulation as in FIG.~10;  Dashed
    curve: Poisson-based approximation, Eq.~\eqref{eq:GFuoss}; Dotted
    curves: augmented maximum entropy model utilizing the first two
    binomial moments, Eq.~\eqref{eq:lnp0}.  The two-moment maximum
    entropy model is qualitatively reliable and therefore gives a
    satisfactory explanation of the observed asymmetry.
\label{fig:MAXENT}}
\end{figure}
\section{Conclusion}
Results for both the [tea][BF$_4$]/PC (FIG.~1) and the ionic liquid
[bmim][BF$_4$] (FIG.~8) identify a natural clustering radius where mean
coordination numbers are near two. This suggests arrangements of the
closest neighbors leading to a structural motif of cation-anion chains
and rings.  In contrast to the atomically detailed [tea][BF$_4$]/PC
results, a corresponding primitive model (FIG.~1) does not display those
clustering signatures (FIG.~6). A generalization (Eq.~\eqref{eq:GFuoss})
of the Fuoss ion-pairing model was obtained by recognizing that the
Poisson distribution is correct when the mean coordination numbers are
low. On the basis of measurable molecular distribution functions, this
generalization also establishes the distribution of  molecular nearest
neighbors for computational analysis of bi-molecular reactive processes
in solution. This Poisson-based model is accurate for the
[tea][BF$_4$]/PC results, both for the primitive model and the
atomically detailed case. For [tea][BF$_4$]/PC, the atomically detailed
numerical results and the statistical model identify
\emph{solvent-separated} nearest-neighbor ion-pairs. Distributions of
nearest-neighbor distances typically depend on which ion of a pair is
taken as the central  ion, \emph{i.e.,} the distribution of  anions
nearest to a cation is different from the distribution of the cations
nearest to an anion. The Poisson-based model is \emph{not} asymmetric in
that way. The numerical data for the ionic liquid [bmim][BF$_4$]
prominently show  the expected asymmetry.  That asymmetry can be treated
by a maximum entropy model based on the expected number of \emph{pairs}
of counter-ions occupying the inner-shell of the central ion,
information extracted from the simulations.

%

\clearpage

\end{document}